\documentclass[10pt, journal]{IEEEtran}

\usepackage{booktabs}
\usepackage{threeparttable}
\usepackage{amsmath,graphicx,cite,amssymb,mathtools}
\usepackage{amsfonts,multirow,bm,array,setspace,stfloats}

\usepackage{float,cite,amssymb}
\usepackage{graphics}
\DeclareGraphicsExtensions{.pdf,.jpeg,.png,.jpg}   
\usepackage{multirow,bm,bbm,array,setspace}
\usepackage{textcomp}
\usepackage{psfrag}
\usepackage{color}
\usepackage{pstricks,enumerate}
\usepackage{bbm}
\usepackage{makecell}
\usepackage{subfigure}

\usepackage{algorithm,algpseudocode}
\usepackage{enumitem}
\usepackage{xpatch}

\makeatletter
\newcommand{\distas}[1]{\mathbin{\overset{#1}{\kern\z@\sim}}}%
\newsavebox{\mybox}\newsavebox{\mysim}
\newcommand{\distras}[1]{%
	\savebox{\mybox}{\hbox{\kern1pt$\scriptstyle#1$\kern1pt}}%
	\savebox{\mysim}{\hbox{$\sim$}}%
	\mathbin{\overset{#1}{\kern\z@\resizebox{\wd\mybox}{\ht\mysim}{$\sim$}}}%
}

\ExplSyntaxOn
\cs_new:Npn \bibColoredItems #1#2
  {
    \clist_map_inline:nn {#2} { \cs_new:cpn {bib@colored@##1} {#1} } 
  }
\ExplSyntaxOff

\newcommand\bib@setcolor[1]{%
  \ifcsname bib@colored@#1\endcsname
    \expandafter\color\expandafter{\csname bib@colored@#1\endcsname}
  \else
    \normalcolor
  \fi
}

\xpatchcmd\@bibitem
  {\item}
  {\bib@setcolor{#1}\item}
  {}{\PatchFailed}

\xpatchcmd\@lbibitem
  {\item}
  {\bib@setcolor{#2}\item}
  {}{\PatchFailed}

\makeatother
\ifCLASSINFOpdf
\else
\fi

\newtheorem{theorem}{Theorem}

\newtheorem{remark}{Remark}

\newcommand{\bX}{\bm{X}}
\newcommand{\bx}{\bm{x}}

\begin{document}
	%
\title{FollowSpot: Enhancing Wireless Communications via Movable Ceiling-Mounted Metasurfaces}
\author{
\IEEEauthorblockN{
    Wenhai Lai, \IEEEmembership{Graduate Student Member,~IEEE},  Kaiming Shen, \IEEEmembership{Senior Member,~IEEE}, \\
    and Rui Zhang, \IEEEmembership{Fellow,~IEEE}
} 
\thanks{
Paper accepted to IEEE Transactions on Communications, February 2026.
This work was supported in part by the National Natural Science Foundation of China under Grant 12426306 and Grant 62331022, and in part by the Guangdong Major Project of Basic and Applied Basic Research (No. 2023B0303000001). \emph{(Corresponding author: Kaiming Shen.)}

Wenhai Lai and Kaiming Shen are with School of Science and Engineering at The Chinese University of Hong Kong (Shenzhen), and Shenzhen Research Institute of Big Data, China (e-mails: wenhailai@link.cuhk.edu.cn;
shenkaiming@cuhk.edu.cn).

Rui Zhang is with the Department of Electrical and Computer Engineering, National University of Singapore, Singapore (e-mail: elezhang@nus.edu.sg).
}
}

%


\maketitle

\begin{abstract}
This work focuses on the optimal placement of metasurfaces (MTSs) onto the ceiling of an industrial manufacturing workshop. In particular, we assume that a total of $M$ MTSs are deployed, and that there are $L$ possible positions for each MTS. The resulting signal-to-noise (SNR) maximization problem is difficult to tackle directly because of the coupling between the placement decisions of the different MTSs. Mathematically, we are faced with a nonlinear discrete optimization problem with $L^M$ possible solutions. A remarkable result shown in this paper is that the above challenging problem can be efficiently solved within $O(ML^2\log(ML))$ time. There are two key steps in developing the proposed algorithm. First, we successfully decouple the placement variables of different MTSs by introducing a continuous auxiliary variable $\mu$; the discrete primal variables are now easy to optimize when $\mu$ is held fixed, but the optimization problem of $\mu$ is nonconvex. Second, we show that the optimization of continuous $\mu$ can be recast into a discrete optimization problem with only $LM$ possible solutions, so the optimal $\mu$ can now be readily obtained. Numerical results show that the proposed algorithm can not only guarantee a global optimum but also reach the optimal solution efficiently.
\end{abstract}

\begin{IEEEkeywords}
Metasurface placement, movable passive antennas, passive beamforming, discrete optimization, global solution.
\end{IEEEkeywords}

\section{Introduction}
\IEEEPARstart{M}{etasurface} (MTS) \cite{al2017recent, Aobo2018MTS}, a.k.a. intelligent reflecting surface \cite{wu2019intelligent,najafi2021physics} or reconfigurable intelligent surface \cite{bjornson2020reconfigurable,huang2019reconfigurable,di2020smart}, is an emerging wireless device that consists of electromagnetic meta-atoms, designed to manipulate the phases and amplitudes of reflected signals to enhance wireless communications. This work focuses on the optimal placement of MTSs onto the ceiling of an industrial manufacturing workshop that has been extensively studied \cite{ren2021intelligent, khan2024performance, liu2022path}. This topic is motivated from three main perspectives:

\begin{itemize}
    \item \emph{Why MTS?} MTS is a passive equipment in the sense that it does not emit any signals itself, so the power consumption can be significantly reduced as compared to the relay and base station, especially considering the large area of the workshop.
    \item \emph{Why movable MTS?} Because the blockages such as automated guided vehicle (AGV) are moving over time in the workshop environment, the shadowing effects and the blind spots are changing dynamically, so it is advisable to relocate MTSs accordingly.
    \item \emph{Why mount MTS to ceiling?} For a typical workshop, there is a large overhead space for the MTS deployment. Moreover, MTS on the ceiling can induce many more line-of-sight (LOS) reflection paths.
\end{itemize}
While most existing works assume that MTSs are placed at the fixed positions, this paper proposes a new paradigm wherein the MTSs can move around flexibly to track the target receiver, referred to as the \emph{wireless follow spot}. If there are $M$ MTSs in total and each MTS can be moved to any of the $L$ possible positions, then the MTS placement problem boils down to a discrete optimization problem with $L^M$ possible solutions, while finding the optimal solution via exhaustive search over them can incur prohibitive complexity when $L$ and $M$ are large. To tackle this fundamental challenge, we show in this work that the global optimum for the single-user case can be attained efficiently within $O(ML^2\log(ML))$ time.

To control the reflected channels, many previous works assume that a phase shifter is deployed at each meta-atom of MTS, so the key problem is then how to coordinate the phase shift variables across different meta-atoms, namely the passive beamforming problem. A variety of optimization techniques have been considered in this area, ranging from the semidefinite relaxation (SDR) \cite{ zheng2021double,xie2020max, yao2023superimposed} to the fractional programming (FP) \cite{feng2020physical,zhang2022active,zhang2021joint}, the successive convex approximation (SCA) \cite{mu2020exploiting, zhao2021two}, the minorization-maximization (MM) \cite{huang2019reconfigurable, shen2020beamforming, yu2024energy}, and the alternating direction method of multipliers (ADMM) \cite{ning2020beamforming, niu2021weighted}. However, when it comes to the large-scale case, the MTS model in the above works becomes costly and prone to malfunction because it requires using phase shifters massively. As such, \cite{chan20153d,xiong2017customizing,2022millimirror} suggest using an MTS without phase shifters, which is also called the reflector, and show that it can already bring significant performance gain so long as the reflector is placed properly. The recent work \cite{Arun2020RFocus} proposes another phase-shifter-free MTS model that controls the reflected channels by tuning the {\footnotesize{ON-OFF}} status of each meta-atom. In contrast, our work considers using multiple phase-shifter-free MTSs to control the reflected channels by adjusting their positions.

The position of MTS plays a crucial role because it directly impacts the phases and magnitudes of reflected channels. The previous works \cite{wang2023average,bai2024intelligent,mu2021joint} assume that the MTS can be placed anywhere within a given area, so the MTS placement problem can be formulated as a continuous optimization problem. Although \cite{wang2023average,bai2024intelligent,mu2021joint} pursue various optimization objectives (i.e., the average rate maximization, the total power minimization, and the weighted sum rate maximization), they all adopt the successive convex approximation (SCA) method to optimize the continuous variables of the MTS positions. However, the placement of MTS can be faced with complicated spatial constraints in practice. To address this issue, the authors of \cite{efrem2023joint} assume that the position of MTS is limited to a set of discrete spots. The proposed method in \cite{efrem2023joint} is to first solve the relaxed continuous problem and then round the solution to the discrete set. This work also imposes the discrete constraint on the position of each MTS. Unlike the heuristic method in \cite{efrem2023joint}, our proposed method in this work guarantees the global optimality.

\begin{figure}[t]
\centering
\includegraphics[width=0.6\linewidth]{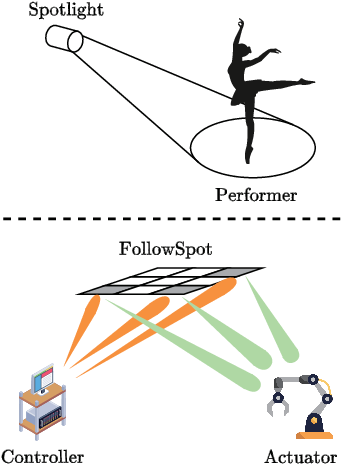}
\caption{Resembling the theatre spotlight, the FollowSpot scheme aims to project the signal beam onto the target receiver (which is the actuator in our case) by placing multiple MTSs properly.}
    \label{fig:followspot}
\end{figure}

Clearly, there would be much more free space if the MTS is placed in the air rather than on the ground, aside from the benefit of providing line-of-sight (LoS) channels. As such, combining the unmanned aerial vehicle (UAV) technology with the MTS has become a research hotspot in recent years. While \cite{lu2021aerial, sun2024aoi, wang2023covert} assume a hovering UAV with the MTS attached, many other works \cite{jiang2024aerial,zhou2024flying,huang2024joint,truong2023flyreflect,chen2025energy} allow the UAV to move around carrying the MTS. However, there are two main issues with the UAV-based MTS: (i) the limited battery life of UAV; (ii) the limited load capacity of UAV. To sidestep the above bottlenecks, another idea \cite{huang2024passive,pan2022sum,gao2023multi,gao2023intelligent,yang2023metaslam} is to install the MTS onto the ceiling if the transmission takes place indoors. Differing from the previous works \cite{huang2024passive,pan2022sum,gao2023multi,gao2023intelligent,yang2023metaslam} that deploy the MTSs at the fixed positions on the ceiling, this work allows the MTS positions to be adjusted over time (under the discrete position constraints), thus enabling a follow-spot effect when the target receiver is moving. It is worth noting that the idea of manipulating channels through spatial positioning stems from another frontier wireless technology called \emph{movable antennas} \cite{zhu2024movable, mei2024movable,Zhuravlev2015,Basbug2017}; it has been shown in \cite{marnat2013} that movable antennas can be realized efficiently by the micro-electro-mechanical systems (MEMS). The system model considered in this paper can be recognized as \emph{movable passive antennas mounted to ceiling}. The present paper aims to eliminate the blind spots from the industrial manufacturing workshop, which are caused by the blockages such as the automated guided vehicle (AGV). Because the large-scale shadowing effect is our main consideration, the latency due to the mechanical movement is tolerable. When it comes to the fast-varying deep fading (e.g., caused by the multipath effect), the physical repositioning cannot keep pace with the channel variation, so it is more advisable to tune the phase shifts of MTSs to combat fading. As a result, the phase shift optimization is performed more frequently than the MTS relocation.


The main contributions of this work are three-fold:
\begin{itemize}
\item We propose a novel paradigm of deploying movable MTSs onto the ceiling. The key idea is to control the reflected channels by adjusting the MTS positions, so that the phase shifts are not required anymore.

\item If there are $M$ MTSs and each MTS has $L$ possible positions, then there are $L^M$ candidate solutions. We propose an efficient discrete optimization algorithm that reaches the global optimum for the single-receiver case within $O(ML^2\log(ML))$ time.
    
\item We further extend the proposed MTS placement algorithm to the multi-receiver case. Simulations show that the extended algorithm outperforms the benchmark methods significantly.
\end{itemize}

\begin{table}
[t]\renewcommand\arraystretch{1.5}
\footnotesize
\centering
\caption{List of Main Variables}
\begin{tabular}{|c||l|}
\hline
\textbf{Symbol} & \textbf{Definition} \\ \hline
\hline
$M$ & number of MTSs \\ \hline
$L$ & number of candidate positions in each zone \\ \hline
$P$ & the transmit signal power \\ \hline
$\sigma^2$ & the noise power \\ \hline
$\mathcal{M}$ & set of all MTSs. \\ \hline
$\mathcal{L}_m$ & set of possible positions for MTS $m$. \\ \hline
$h_{0}$ & direct channel from controller to actuator \\ \hline
$h_{m,\ell}$ & reflected channel induced by MTS $m$ at position $\ell$ \\ \hline
$x_{m,\ell}$ & it is 1 if MTS $m$ is at position $\ell$, and 0 otherwise \\ \hline
$\bX$ & an $M\times L$ matrix whose $(m,\ell)$th entry is $x_{m,\ell}$ \\ \hline
$\bx_m$ & $m$th row vector of $\bX$\\ \hline
$\mu$ & an auxiliary variable which is restricted to the unit circle\\ \hline
$\mathcal A_{m,\ell}$ & an sector of unit circle used to optimize $\bX$ \\ \hline
$\xi_\ell$ & a point on the unit circle used to decide $\mathcal A_{m,\ell}$\\ 
\hline
\end{tabular}
\label{tab:notation}
\end{table}

The rest of the paper is organized as follows. Section \ref{sec:sys} describes the system model and problem formulation. Section \ref{sec:opt_placement}  proposes a novel MTS-placement algorithm; its extension to the multi-receiver case is discussed in Section \ref{sec:multiple}. Section \ref{sec:simulation} shows the numerical results. Finally, Section \ref{sec:conclusion} concludes the paper. Throughout the paper, we denote by $\mathfrak{Re}\{u\}$ the real part of a complex number $u$, and $\angle u$ the phase. A summary of the main variables used in the paper is provided in Table \ref{tab:notation}.

\section{System Model}
\label{sec:sys}
\begin{figure*}
    \centering
    \includegraphics[width=0.6\linewidth]{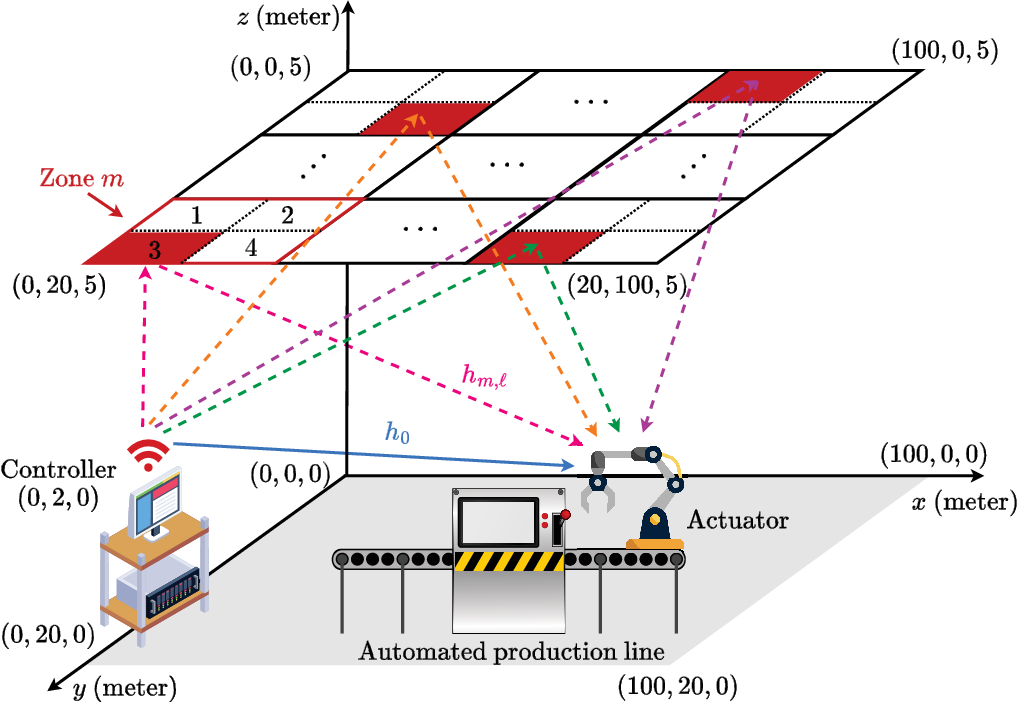}
    \caption{In this example, the ceiling is divided into multiple square zones. Each MTS can move around within its zone; each small square area of the zone represents a possible position of the MTS, and the current position of each MTS is highlighted in red. For instance, for a particular MTS $m$, its four possible positions are indexed by $\{1,2,3,4\}$, and it is currently placed at position $3$. This model is considered later in simulations in Section \ref{sec:simulation}.}
    \label{fig:IRS_system}
\end{figure*}

We consider the wireless transmission in an industrial manufacturing workshop in which a controller sends a stream of control data to an actuator of the automated production line; the multi-actuator case will be discussed later in Section \ref{sec:multiple}. For the controller-to-actuator communication, this work suggests a new paradigm called \emph{FollowSpot}.
By this paradigm, the ceiling of the industrial manufacturing workshop is divided into multiple zones, with a phase-shifter-free MTS mounted onto each zone. The controller coordinates the MTS by the wireline connections to ensure robust low-latency transmission. Moreover, the size of the zone of each MTS should be limited so as to reduce the delay of mechanic motion. Unlike the conventional MTSs that rely on phase shifters, the MTSs in our case manipulate the reflected channels through their placement. Moreover, we assume that there is a discrete set of possible positions for the MTS within each zone\footnote{Indeed, the continuous degrees of freedom can bring additional performance gain, but at the cost of much more overhead. In that case, we must build a channel model that can accurately predict the CSI at an arbitrary position, which not only is time-consuming but also incurs large CSI error. In contrast, when each MTS is limited to a discrete set of candidate positions, we just need to measure CSI at these grids, so the CSI acquisition becomes much easier. Moreover, when each MTS is allowed to be continuously placed, the control message requires significantly more bits.}. As such, the proposed \emph{FollowSpot} aims to coordinate the positions of the multiple MTSs to focus the reflected beam on the actuator, as shown in Fig.~\ref{fig:followspot}.

We denote by $M$ the total number of zones. Each zone has $L$ candidate positions for its associated MTS. The set of all zones is $\mathcal{M}=\{1,2,\ldots, M\}$ and the set of all candidate positions for zone $m$ is $\mathcal{L}_m=\{1,2,\ldots, L\}$. The MTS is indexed as $m$ if it is mounted onto the zone $m\in\mathcal{M}$. Assume that each MTS consists of $N$ meta-atoms. Let $h_0\in\mathbb{C}$ be the direct channel from the controller to the actuator. When the MTS $m$ is placed at position $\ell\in\mathcal{L}_m$, let $f_{m,n,\ell}\in\mathbb{C}$ be the channel from the controller to the $n$th meta-atom of MTS $m$, and let $g_{m,n, \ell}\in\mathbb{C}$ be the channel from the $n$th meta-atom of MTS $m$ to the actuator. Thus, the reflected channel $h_{m,n,\ell}\in\mathbb C$ induced by the $n$th meta-atom of MTS $m$ with respect to position $\ell$ is given by
\begin{equation}
\label{hn:sub}
    h_{m,n,\ell} = f_{m,n,\ell}\times g_{m,n, \ell}.
\end{equation}
The overall reflected channel $h_{m,\ell}\in\mathbb C$ induced by MTS $m$, when deployed at position $\ell$, amounts to the superposition of reflected channels across all different meta-atoms:
\begin{equation}
\label{hn}
    h_{m,\ell} = \sum_{n=1}^N h_{m,n,\ell}.
\end{equation}
In this work, we assume that the channel state information (CSI) $h_0$ and all $\{h_{m, \ell}\}$ are known \emph{a priori}\footnote{Notice that our method requires accurate CSI. But more importantly, since the reflection paths from the ceiling are line-of-sight (LOS)---which is the main motivation of using the ceiling-mounted MTS, it suffices to measure the channel for one meta-atom, with the channels of the rest meta-atoms recovered by the uniform planner array (UPA). Thus, the CSI measurement is scalable. Furthermore, the MTS placement aims to eliminate blind spots, so its input CSI is the large-scale fading component. Since the motion of the obstacles (e.g., AGV) is predictable in the workshop scenario, the large-scale CSI acquisition is not that difficult. In addition, to combat the small-scale fading, the MTS relocation
is not enough, and it further requires the phase shift optimization based on the small-scale CSI.}. Note that we do not assume any particular channel models for $f_{m,n,\ell}$ and $g_{m,n,\ell}$.

Let $x_{m,\ell}\in\{0,1\}$ be a binary placement variable denoting whether or not the MTS $m$ is deployed at position $\ell$. Moreover, we require $\sum_{\ell\in\mathcal{L}_m} x_{m,\ell}=1,\,\forall m$ to ensure that only a single MTS is mounted onto each zone. For the transmitted symbol $s$ and the background noise $z\sim\mathcal{CN}(0,\sigma^2)$, the received signal $y\in\mathbb{C}$ at the actuator is given by
\begin{equation}
    y = \left(h_0+\sum_{m\in\mathcal{M}}\sum_{\ell\in\mathcal{L}_m} x_{m,\ell}h_{m,\ell}\right)s + z.
\end{equation}
Then, the signal-to-noise ratio (SNR) of the actuator is
\begin{equation}
    \mathrm{SNR} = \frac{P}{\sigma^2}\left|h_0+\sum_{m\in\mathcal{M}}\sum_{\ell\in\mathcal{L}_m} x_{m,\ell}h_{m,\ell}\right|^2,
\end{equation}
where $P$ is the transmit power.

We seek the optimal placement of MTSs to maximize the SNR. Moreover, it turns out that maximizing the received SNR amounts to maximizing the overall channel strength. Thus, the placement optimization problem can be written as a nonlinear integer optimization problem:
\begin{subequations}
\label{placement_problem}
\begin{align}
    \underset{\bX}{\text{maximize}} \quad &f(\bX)\coloneqq\left|h_0+\sum_{m\in\mathcal{M}}\sum_{\ell\in\mathcal{L}_m} x_{m,\ell}h_{m,\ell}\right|
    \label{placement_problem:obj}\\
    \text {subject to} \quad &x_{m,\ell}\in \{0,1\},\,\forall m,\forall\ell, \\
    &\sum_{\ell\in\mathcal{L}_m} x_{m,\ell} = 1,\,\forall m, \label{placement_problem:cons2}
\end{align}
\end{subequations}
where $\bX=[x_{m,\ell}]$. Moreover, in the rest of the paper, we use $\bx_m$ to denote the $m$th row vector of $\bX$, i.e., $\bx_m=[x_{m,1},x_{m,2},\ldots,x_{m,L}]$, which corresponds to the placement decision of MTS $m$. The above problem is difficult to tackle directly because of the discrete constraint and the coupling between distinct $x_{m,\ell}$ and $x_{m,\ell^\prime}$. Some previous works \cite{hammer2002maximizing,nie2016linear} even thought such type of problem is NP-hard. We remark that the exhaustive search is impractical because there are a total of $L^M$ possible solutions. As a striking result shown in this work, the global optimum of problem \eqref{placement_problem} can be reached within $O(ML^2\log(ML))$ time.

\begin{remark}
    We would like to clarify that the MTS placement coordination aims to eliminate the blind spots caused by the large-scale shadowing effect, so it suffices to consider a single-antenna channel. To further reap the gain of MIMO transmission, we need the refined CSI that captures the multipath effect. Thus, after the positions of MTSs have been determined, we can refine the CSI measure and then perform any existing beamforming method to enhance the MIMO transmission.
\end{remark}

\section{Proposed MTS-Placement Algorithm}
\label{sec:opt_placement}

We pursue a polynomial-time algorithm that guarantees the global optimum of problem \eqref{placement_problem}. Before proceeding to that, we first rewrite problem \eqref{placement_problem} by introducing a continuous auxiliary variable, so as to linearize the objective function and enable a divide-and-conquer strategy.

\subsection{Problem Reformulation}

We propose bounding the objective \eqref{placement_problem:obj} from below as
\begin{align}
    \label{eq:rewrite_fx}
    f(\bX) &= \left|h_0+\sum_{m\in\mathcal{M}}\sum_{\ell\in\mathcal{L}_m} x_{m,\ell}h_{m,\ell}\right| \notag \\
    &\overset{(a)}{=}\left|\mu\left(h_0+\sum_{m\in\mathcal{M}}\sum_{\ell\in\mathcal{L}_m} x_{m,\ell}h_{m,\ell}\right)\right| \notag \\
    &\overset{(b)}{\geq} \mathfrak{Re}\left\{\mu\left(h_0+ \sum_{m\in\mathcal{M}}\sum_{\ell\in\mathcal{L}_m} x_{m,\ell}h_{m,\ell}\right)\right\},
\end{align}
where $(a)$ follows by introducing a normalized complex number $\mu\in\mathbb C$ with $|\mu|=1$, and $(b)$ follows as $\mathfrak{Re}(z)\le |z|$ for any complex number $z$. 
We remark that the above lower bound is tight since the equality in $(b)$ holds whenever
\begin{equation}
    \label{eq:angle_equal}
    \angle \mu=-\angle \left(h_0+\sum_{m\in\mathcal{M}}\sum_{\ell\in\mathcal{L}_m} x_{m,\ell}h_{m,\ell}\right).
\end{equation}
Thus, letting
\begin{equation}
\label{func g}
    g(\bX,\mu)
    = \mathfrak{Re}\left\{\mu\left(h_0+ \sum_{m\in\mathcal{M}}\sum_{\ell\in\mathcal{L}_m} x_{m,\ell}h_{m,\ell}\right)\right\},
\end{equation}
problem \eqref{placement_problem} is equivalent to
\begin{subequations}
\label{max_max}
\begin{align}
    \underset{\bX,\,\mu}{\text{maximize}}\quad &  g(\bX,\mu) \\
    \text {subject to} \quad &x_{m,\ell}\in\{0,1\},\,\forall m,\forall \ell, \label{max_max:cons1}\\
    &\sum_{\ell\in\mathcal{L}_m} x_{m,\ell} = 1,\,\forall m, \label{max_max:cons2}\\
    &|\mu|=1.
\end{align}
\end{subequations}
In problem \eqref{max_max}, assuming that the optimal solution of $\mu$, denoted by $\mu^\star$, is already known, we then focus on optimizing $\bX$. The key observation is that problem \eqref{max_max} can be divided into a set of subproblems on a per-MTS basis, each only depending on a particular $\bx_m$:
\begin{subequations}
    \begin{align}
    \underset{\bx_m}{\text{maximize}}\quad&\mathfrak{Re}\left\{\mu^\star\sum_{\ell\in\mathcal{L}_m} x_{m,\ell}h_{m,\ell}\right\} \\
    \text {subject to} \quad &x_{m,\ell}\in\{0,1\},\forall \ell, \\
    &\sum_{\ell\in\mathcal{L}_m} x_{m,\ell} = 1.
    \end{align}
\end{subequations}
Clearly, each $\bx_m$ can be optimally determined as
\begin{equation}
    \label{eq:x_ml}
    x^\star_{m,\ell}=\begin{cases}
  1, & \text{if } \ell=\ell_m^\star, \\
  0,  & \text{otherwise},
\end{cases}
\end{equation}
where $\ell_m^\star$ is given by
\begin{equation}
\label{opt ell}
\ell_m^\star = \arg\max_{\ell\in\mathcal{L}_m}\,\mathfrak{Re}\{\mu^\star h_{m,\ell}\}.
\end{equation}

Recall that it is difficult to decide $\bX$ in the original problem \eqref{placement_problem} because of the coupling between the different $\bx_m$'s in the nonlinear objective function. But now, thanks to the reformulation as in \eqref{max_max}, the new objective $g(\bX,\mu)$ is a linear combination of the different $\bx_m$'s (so long as $\mu$ is fixed) and thus allows efficient solving of $\bX$. We can now treat the solution of $\bX$ as a function of $\mu$, written $\bX^\star(\mu)$, where each row vector is computed as in \eqref{eq:x_ml}. As a consequence, the joint optimization of $(\bX,\mu)$ in \eqref{max_max} now boils down to optimizing $\mu$ alone:
\begin{equation}
    \label{max_max_final}
    \underset{\mu:\;|\mu|=1}{\text{maximize}}\quad g(\bX^\star(\mu), \mu).
\end{equation}
which is dealt with in the next subsection.

\subsection{Global Solution in Polynomial Time}

Our current goal is to find the optimal $\mu$ in problem \eqref{max_max_final} and then recover the optimal $\bX$ in the original problem \eqref{placement_problem} through \eqref{eq:x_ml}. Denote the unit circle on the complex plane by $\mathcal{C}$, i.e.,
\begin{equation}
    \mathcal C = \{c\in\mathbb C: |c|=1\}.
\end{equation}
Thus, problem \eqref{max_max_final} seeks the optimal point $\mu^\star\in\mathcal C$ to maximize $g(\cdot)$. A naive idea is to search through the unit circle continuously and exhaustively for the optimal $\mu$ in \eqref{max_max_final}. In contrast, as the key result in this subsection, we show that it suffices to consider a finite discrete set of points on the unit circle.


\begin{figure*}[t]
\centering
\subfigure[MTS 1]{
\label{example_MTS1}
\includegraphics[width=0.3\linewidth]{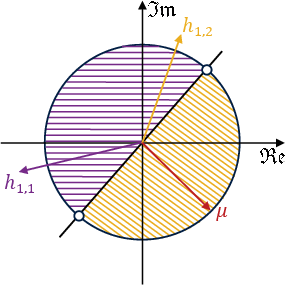}}
\quad
\subfigure[MTS 2]{
\label{example_MTS2}
\includegraphics[width=0.3\linewidth]{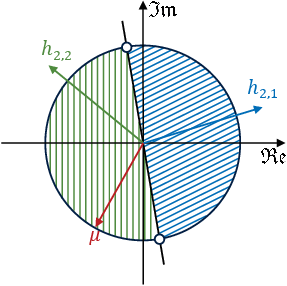}}
\quad
\subfigure[Combined]{
\label{example_combine}
\includegraphics[width=0.3\linewidth]{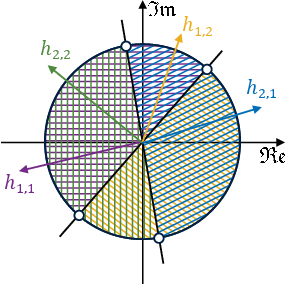}}
\caption{Illustration of the proposed algorithm. Consider two MTSs each with 2 possible positions. (a): The optimal position of MTS 1 depends on the unit vector $\mu$; if $\mu$ lies in the yellow sector, $h_{1,2}$ leads to a larger $f(\bX)$ so we should place MTS 1 at position 2; if $\mu$ lies in the purple sector, $h_{1,1}$ is better so we should place MTS 1 at position 1. (b): Similarly, if $\mu$ lies in the blue sector, we should place MTS 2 at position 1; if $\mu$ lies in the green sector, we should place MTS 2 at position 2. (c): Now we combine the sectorizations of MTSs 1 and 2 to obtain a total of 4 sectors; we just try out each sector for $\mu$ to see which yields the largest $f(\bX)$, thereby obtaining the global optimum. Most importantly, for the general case, the total number of sectors after combination is $O(ML)$, so the proposed algorithm is scalable.}
\label{fig:alg}
\end{figure*}

\begin{figure*}[t]
\centering
\subfigure[]{
\label{example_fake_two}
\includegraphics[width=0.3\linewidth]{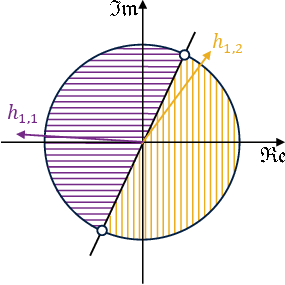}}
\quad
\subfigure[]{
\label{example_fake_three}
\includegraphics[width=0.3\linewidth]{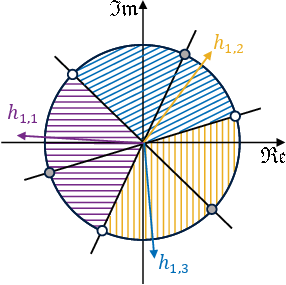}}
\quad
\subfigure[]{
\label{example_fake_four}
\includegraphics[width=0.3\linewidth]{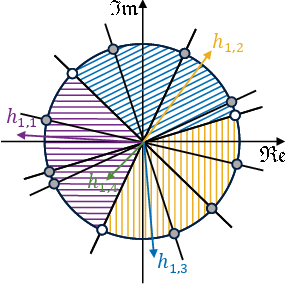}}
\caption{Illustration of the ``fake'' transition point.  Recall that the optimal positions depend on which sector $\mu$ lies in. The sectors are formed by a set of straight-line cuts. Each cut passes through the origin and intersects with the unit circle at two positions; an intersection is referred to as a transition point if the optimal position of MTS changes when $\mu$ passes through it. (a): When $L=2$, there is one straight-line cut, and the two intersections are both transition points. (b): When $L=3$, there are three straight-line cuts and hence six intersections; however, three intersections (marked in gray) are between two sectors that lead to the same optimal position of MTS 1, so the optimal position does not change when $\mu$ passes through any of them; these gray intersections are referred to as ``fake'' transition points. (c): When $L=4$, there are six straight-line cuts and thus twelve intersections; observe that nine of them are fake transition points. For the general case, there are $ML(L-1)/2$ straight-line cuts and $ML(L-1)$ intersections, but the number of fake transition points depends on the specific case.}
\label{fig:example_fake}
\end{figure*}

Let $\mu$ move along the unit circle. The effect of multiplying $\mu$ to $h_{m,\ell}$ can be thought of as rotating the vector of $h_{m,\ell}$ in the complex plane by an angle of $\angle \mu$ counterclockwise. Thus, for the current $\mu$, the solution in \eqref{opt ell} boils down to first rotating all the vectors of $\{h_{m,1},\ldots,h_{m,L}\}$ simultaneously by an angle of $\angle \mu$ counterclockwise and then finding which rotated vector leads to the largest projection on the real axis. Furthermore, note that the ``winner'' vector that achieves the largest projection remains the same $h_{m,\ell}$ when $\mu$ does not move too far. To be more specific, all those unit vectors $\mu$ giving the same solution in \eqref{opt ell} would lie within a sector of the unit disk on the complex plane, as shown in Fig. \ref{fig:alg}. In other words, the solution $\ell_m^\star$ in \eqref{opt ell} equals $\ell'$ if and only if the current $\mu$ lies in the sector associated with $\ell'$, denoted by $\mathcal A_{m,\ell}$. We remark that some $h_{m,\ell}$ may not have its sector, i.e., $h_{m,\ell}$ can never be the ``winner'' vector regardless of the value of $\mu$; this subtle point is illustrated in Fig.~\ref{example_fake_four}. As a consequence, we conclude that there are at most $L$ sectors to consider for MTS $m$.

Moreover, we observe from \eqref{eq:x_ml} that the optimal solution of $\bx_m$ only depends on $\ell^\star_m$; so, equivalently, we can say that the optimal solution of $\bx_m$ only depends on which sector $\mathcal A_{m,\ell}$ contains $\mu^\star$. Most importantly, since there are at most $L$ sectors, we can try putting $\mu$ in each of them and then pick the best one that yields the largest value of $g(\cdot)$ in \eqref{func g}. Thus, it just requires a linear complexity to obtain the optimal $\mu^\star$ and $\bX^\star$. To summarize, we now know how to solve problem \eqref{placement_problem} efficiently when the sectors $\mathcal A_{m,\ell}$ are available. But how do we obtain the sectors $\mathcal A_{m,\ell}$?

Instead of computing each $\mathcal A_{m,\ell}$ directly, we consider the \emph{transition points} between the sectors on the unit circle, i.e., the solution of $\bx_m$ in \eqref{eq:x_ml} alters whenever $\mu$ passes through a transition point, as shown in Fig.~\ref{fig:alg}. Clearly, the transition point $\xi\in\mathcal C$ between the sectors of $h_{m,\ell}$ and $h_{m,\ell'}$ must satisfy the following equation:
\begin{equation}
\label{xi:equation}
\mathfrak{Re}\{\xi h_{m,\ell}\} = \mathfrak{Re}\{\xi h_{m,\ell'}\}.
\end{equation}
Actually, the above equation has exactly two solutions:
\begin{equation}
\label{xi:sol1}
    \xi = e^{\mathrm{j}\big(\frac{\pi}{2}-\angle (h_{m,\ell}-h_{m,\ell^\prime})\big)}
\end{equation}
and
\begin{equation}
\label{xi:sol2}
    \xi = e^{\mathrm{j}\big(\frac{3\pi}{2}-\angle (h_{m,\ell}-h_{m,\ell^\prime})\big)}.
\end{equation}
We then consider equation \eqref{xi:equation} for all $m\in\mathcal{M}$ and all pairs of distinct $\ell,\ell'\in\mathcal{L}_m$. As a consequence, there are $ML(L-1)/2$ such equations, with $ML(L-1)$ solutions. The resulting solutions are sorted counterclockwise (i.e., in an increasing order of their angles) as
$$
\angle\xi_1\le\angle\xi_2\le\ldots\le\angle\xi_{ML(L-1)}.
$$
It is worth noting that equation \eqref{xi:equation} is only a necessary condition for the transition point. In other words, every transition point must appear in the above sequence $(\xi)^{ML(L-1)}_{\ell=1}$, whereas not every $\xi$ is guaranteed to be a transition point. Fig.~\ref{fig:example_fake} provides an example in this regard. The key observation is that the sequence $(\xi_\ell)^{ML(L-1)}_{\ell=1}$ can be efficiently obtained within $O(ML^2)$ time; moreover, blending ``fake'' transition points into the real ones does not prevent us from solving problem \eqref{max_max_final} correctly. As such, we use \eqref{xi:equation} to obtain a superset of the transition points.

\begin{algorithm}[t]
\caption{MTS placement for single-user transmission}
\label{alg:OP}
\begin{algorithmic}[1]
    \For{$m\in\mathcal{M}$}
    \State Find the solutions $\xi_\ell$ to equation \eqref{xi:equation} in closed form as in \eqref{xi:sol1} and \eqref{xi:sol2} for all pairs of distinct $\ell,\ell'\in\mathcal{L}_m$.
    \EndFor
    \State Sort the solutions $\xi_\ell$ counterclockwise.
    \For{$k=1,2,\ldots,ML(L-1)$}
    \State Put $\mu$ inside the $\xi_k$-sector.
    \State Compute each $\ell_m^\star$ according to \eqref{opt ell}.
    \State Compute each $\bx_m^\star$ according to \eqref{eq:x_ml}.
    \State Compute the resulting value $f(\bX^\star)$.
    \EndFor
    \State Find out which $\xi$-sector yields the largest value of $f(\bX^\star)$, and output its corresponding $\bX^\star$.
\end{algorithmic}
\end{algorithm} 

We now plot the sequence $(\xi_\ell)^{ML(L-1)}_{\ell=1}$ on the unit circle as in Fig.~\ref{fig:example_fake}. These points divide the unit disk into at most\footnote{Note that $\xi_1,\xi_2,\ldots,\xi_{ML(L-1)}$ may not be distinct.} $ML(L-1)$ sectors; these sectors are referred to as the \emph{$\xi$-sector}. Because $\{\xi_\ell\}$ is a superset of the transition point set, each sector $\mathcal A_{m,\ell}$ must be composed of one or more $\xi$-sector. Intuitively, the decomposition of the unit disk by $\xi_\ell$ is finer than that by the transition points. In particular, it is easy to see that when $\mu$ moves around within the same $\xi$-sector, the solution of $\bx_m$ remains the same, just like the aforementioned case of $\mathcal A_{m,\ell}$. Then we just try putting $\mu$ inside each $\xi$-sector and find out which $\xi$-sector yields the best $\bX^\star$---which must be the actual solution to the original problem \eqref{placement_problem}. Regarding the complexity, the most time-consuming part is to sort out the sequence $(\xi)^{ML(L-1)}_{\ell=1}$, so the overall complexity equals $O(ML^2\log(ML^2))$, which further simplifies into $O(ML^2\log(ML))$. Algorithm \ref{alg:OP} summarizes all the details.

\begin{theorem}
    Algorithm \ref{alg:OP} yields a globally optimal solution to problem \eqref{placement_problem} within $O(ML^2\log(ML))$ time.
    \label{theorem:opt}
\end{theorem}

We have shown that Algorithm \ref{alg:OP} can recover the corresponding $\bX^\star$ given a particular $\mu$. Since the optimal $\mu$ must reside in one of the different $\xi$-sectors, the solution produced by Algorithm \ref{alg:OP} must be optimal for problem \eqref{max_max_final}. The above theorem immediately follows by the equivalence between problem \eqref{placement_problem} and problem \eqref{max_max_final}.

\section{Extension to Multi-Receiver Case}
\label{sec:multiple}

The goal of this section is to extend Algorithm \ref{alg:OP} to multiple actuators. We assume that the controller broadcasts a common message to $U\ge2$ actuators. If one wishes to deliver separate messages to multiple actuators, we propose using the so-called ``Occupy CoW'' scheme \cite{swamy2017real,liu2018d2d,ayoughi2019interference,chang2019optimizing} (which was proposed exactly for the industrial communications) to somehow convert the separate-message broadcast to the common-message broadcast. Denote by $h_{0,u}$ the direct channel from the controller to the $u$th actuator, and $h_{m,\ell,u}$ the reflected channel induced by MTS $m$ for the $u$th actuator when the MTS is placed at position $\ell$ in its zone, where $m\in\mathcal{M}$, $\ell\in\mathcal{L}_m$, and $u=1,\ldots,U$. The resulting SNR of actuator $u$ is then given by
\begin{equation}
    \mathrm{SNR}_u = \frac{P}{\sigma^2}\left|h_{0,u}+\sum_{m\in\mathcal{M}}\sum_{\ell\in\mathcal{L}_m} x_{m,\ell}h_{m,\ell, u}\right|^2.
\end{equation}
Thus, the placement of each MTS $m$, $\bx_m$, now impacts the performance of all the $U$ actuators.

We seek the optimal $\bX$ to maximize the worst SNR among the actuators:
\begin{subequations}
\label{MU_problem}
\begin{align}
    \underset{\bX}{\text{maximize}} \quad &\min_u \{\mathrm{SNR}_u\}
    \\
    \text {subject to} \quad &x_{m,\ell}\in \{0,1\},\,\forall m,\forall\ell, \\
    &\sum_{\ell\in\mathcal{L}_m} x_{m,\ell} = 1,\,\forall m.
\end{align}
\end{subequations}

We propose a natural extension of Algorithm \ref{alg:OP} based on majority voting \cite{xu2024blind,lam1997application} to solve problem \eqref{MU_problem} sub-optimally. First, run Algorithm \ref{alg:OP} for each user $u$ separately. Let $\bX^{(u)}$ be the solution of Algorithm \ref{alg:OP} when actuator $u$ is the target user in consideration. As before, let $\bx^{(u)}_m$ be the $m$th row vector of $\bX^{(u)}$, and let $x^{(u)}_{m,\ell}$ be the $\ell$th entry of $\bx^{(u)}_m$. Clearly, $\bX^{(u)}$ obtained for different actuators are usually not equal. We now adopt a \emph{majority voting} strategy to reach a consensus. Specifically, for each $\bx_m$ (which represents the placement of MTS $m$), we collect the decisions of $\bx^{(u)}_m$ from the different $u$, and place the MTS at the position that receives the most votes:
\begin{equation}
    \label{eq:voting_position}
    x_{m,\ell}=\begin{cases}
  1, & \text{if } \ell=\ell_m^\star, \\
  0,  & \text{otherwise},
\end{cases}
\end{equation}
where
\begin{equation}
    \label{eq:compute_ell_MU}
    \ell_m^\star = \arg\max_{\ell\in\mathcal{L}_m}\,\sum^U_{u=1}x^{(u)}_{m,\ell}.
\end{equation}
The above steps are summarized in Algorithm \ref{alg:MV}. It is evident that the overall complexity is $O(UML^2\log(ML))$.

\begin{remark}
    The proposed MTS placement method can be readily combined with the existing algorithms for phase shifting and power transmission in an alternating fashion. Specifically, after the positions of MTSs have been determined, we can refine the CSI measure and then perform any existing algorithms for phase shifting and power transmission.
\end{remark}

\begin{algorithm}[t]
\caption{MTS placement for multi-user broadcast}
\label{alg:MV}
\begin{algorithmic}[1]
    \For{$u=1,2,\ldots,U$}
    \State Run Algorithm \ref{alg:OP} for actuator $u$.
    \EndFor
    \For{$m\in\mathcal{M}$}
    \State Compute each $\ell_m^\star$ according to \eqref{eq:compute_ell_MU}.
    \State Compute each $\bx_m$ according to \eqref{eq:voting_position}.
    \EndFor
    \State Output the optimized $\bX$.
\end{algorithmic}
\end{algorithm}

\section{Numerical Results}
\label{sec:simulation}

We now numerically evaluate the performance of the proposed algorithms. Consider a $100 \times 20$ ceiling in meters, which is divided into $M=M_x\times M_y$ equal rectangular zones as shown in Fig. \ref{fig:IRS_system}. Each zone provides $L$ candidate positions for its MTS. The ceiling height is $5$ meters. The actuator(s) are randomly distributed on the ground. We set $M=6\times 5=30$, $L=6$, and $N=100$ by default.

We then specify the channel model in our simulations. Each of the channels $\{h_0,f_{m,n,\ell}, g_{m,n, \ell}\}$ is modeled as Rician fading:
\begin{equation}
h=\sqrt{\gamma}\left(\sqrt{\frac{\delta}{1+\delta}}\overline{h}+\sqrt{\frac{1}{1+\delta}}\widetilde{h}\right), 
\end{equation}
where $0<\gamma<1$ is the pathloss factor, $
\delta>0$ is the Rician factor, $\overline{h}\in\mathbb C$ is the normalized line-of-sight (LoS) component, and the fading component $\widetilde{h}$ is a random variable drawn from the complex Gaussian distribution $\mathcal{CN}(0,1)$. The parameter of the pathloss follow \cite{ren2022configuring, Xu2024Coordinating,2021JiangtJSAC}, which is generated according to
\begin{equation}
    \gamma = 10^{-(30+22\log_{10}(d))/10},
\end{equation}
where $d$ is the corresponding distance in meters. The LoS components are generated according to $\overline{h}=\exp(-\mathrm{j}2\pi d/\lambda)$ where $\lambda\approx10$ cm is the wavelength when the carrier frequency equals 2.6 GHz. Regarding the Rician factor, we set $\delta=15$ for all channels.

\begin{figure*}[t]
\centering
\begin{minipage}{0.49\linewidth}
    \centering
    \includegraphics[width=\linewidth]{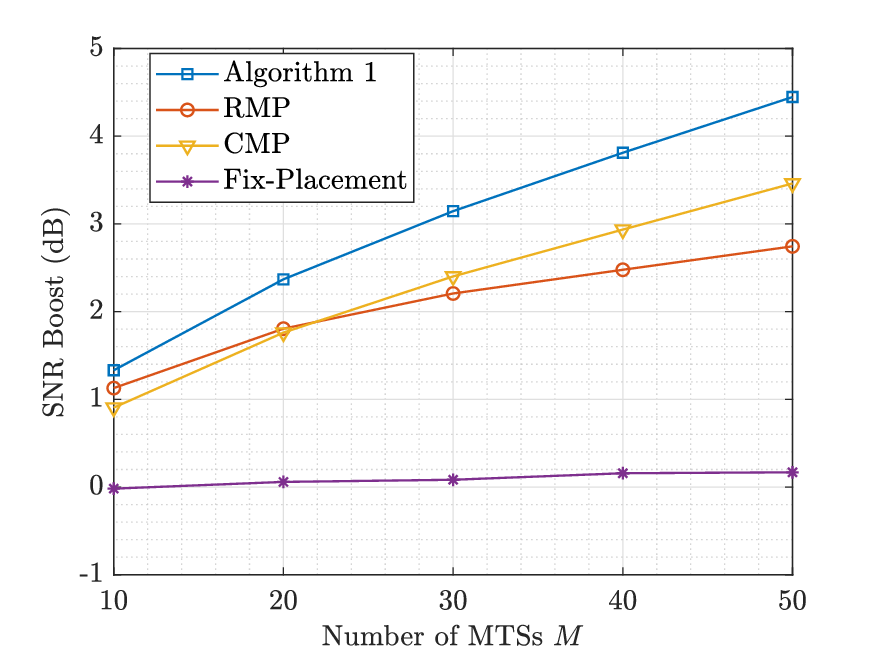}
    \caption{SNR boost vs. the number of MTSs $M$.}
    \label{fig:SNR_vs_M}
\end{minipage}
\begin{minipage}{0.49\linewidth}
    \centering
    \includegraphics[width=\linewidth]{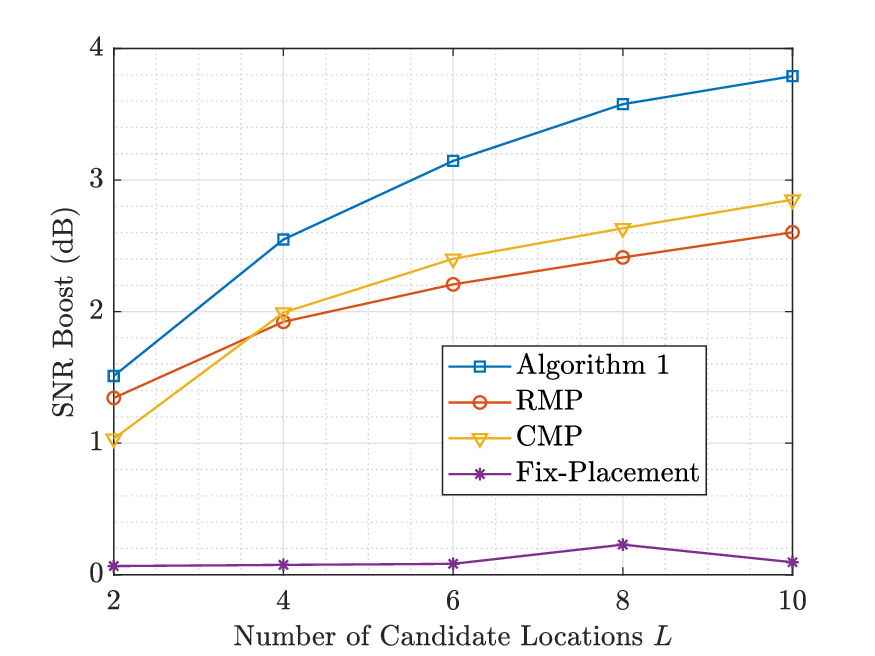}
    \caption{SNR boost vs. the number of candidate positions $L$.}
    \label{fig:SNR_vs_L}
\end{minipage}

\begin{minipage}{0.49\linewidth}
    \centering
    \includegraphics[width=\linewidth]{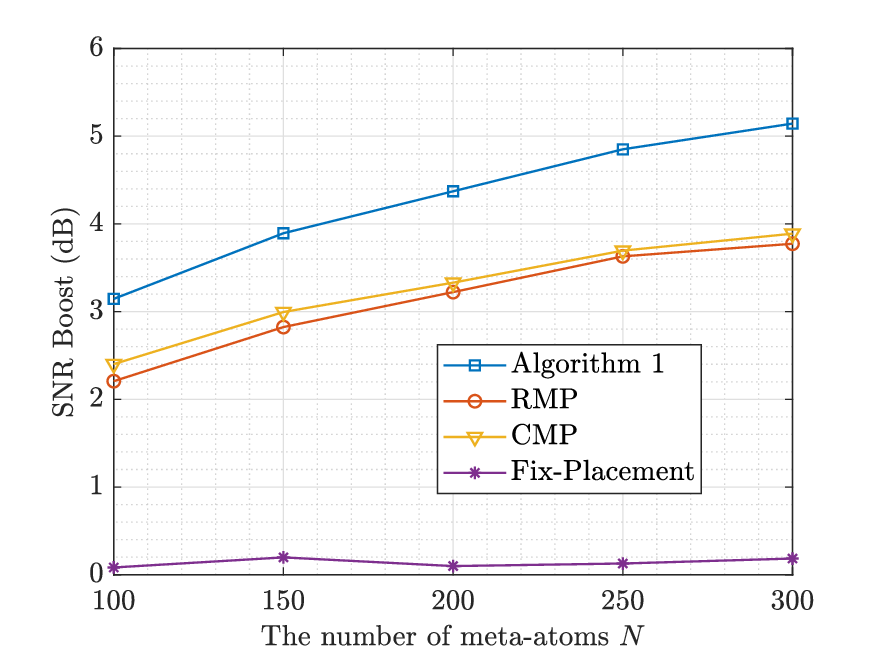}
    \caption{SNR boost vs. the number of meta-atoms $N$.}
    \label{fig:SNR_vs_N}
\end{minipage}
\begin{minipage}{0.49\linewidth}
    \centering
    \includegraphics[width=\linewidth]{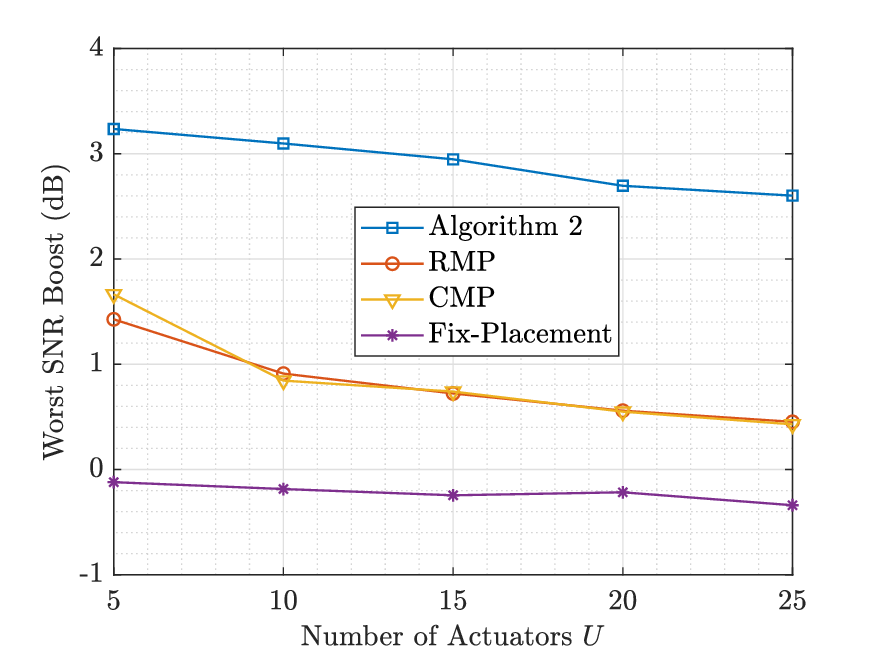}
    \caption{Worst SNR boost vs. the number of actuators $U$.}
    \label{fig:SNR_vs_U}
\end{minipage}
\end{figure*}

The rest of the simulation parameters are set as follows unless otherwise stated. The transmit power level $P=30$ dBm and the background noise power level $\sigma^2=-80$ dBm. All simulation results are generated by averaging out 1000 realizations of fading channels. As for the simulation platform, our simulations are conducted on a computer with a 4.6 GHz i5-11500 CPU and 16 GB RAM.

We compare the proposed algorithms with the following benchmarks:
\begin{itemize}
    \item \emph{Channel Matching Placement (CMP) \cite{wang2020dynamic}:} For the single-receiver case, it involves deploying the MTS $m$ to align the induced reflected channel with the direct channel $h_0$ as closely as possible. For the multi-receiver case, run the single-receiver CMP method for each actuator and then choose the best among the $U$ solutions.

    \item \emph{Random-Max Placement (RMP) \cite{you2020fast}:} Randomly generate $M\times L$ different $\bX$ and choose the best for maximizing the SNR in the single-receiver case and the worst SNR in the multi-receiver case, respectively. 

    \item \emph{Fix-Placement}: Fix the randomly assigned positions of the $M$ MTSs.
\end{itemize}
We compare the performance of the proposed Algorithm \ref{alg:OP} with the above methods in terms of the SNR boost at the actuator:
\begin{equation*}
    \text{SNR boost} = \frac{\text{SNR with MTSs deployed}}{\text{SNR without MTS}}, 
\end{equation*}
and compare the proposed Algorithm \ref{alg:MV} with benchmarks in terms of the worst SNR boost defined similarly.

\begin{figure*}[t]
\centering
\begin{minipage}{0.49\linewidth}
    \centering
    \includegraphics[width=\linewidth]{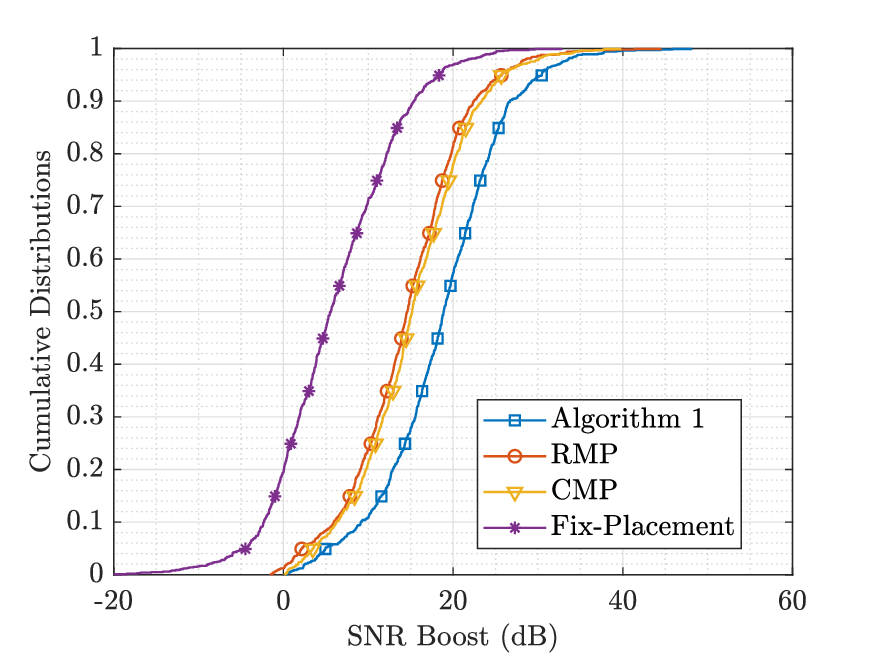}
    \caption{CDF of the SNR boost in the NLoS case.}
    \label{fig:NLOS_SU}
\end{minipage}
\begin{minipage}{0.49\linewidth}
    \centering
    \includegraphics[width=\linewidth]{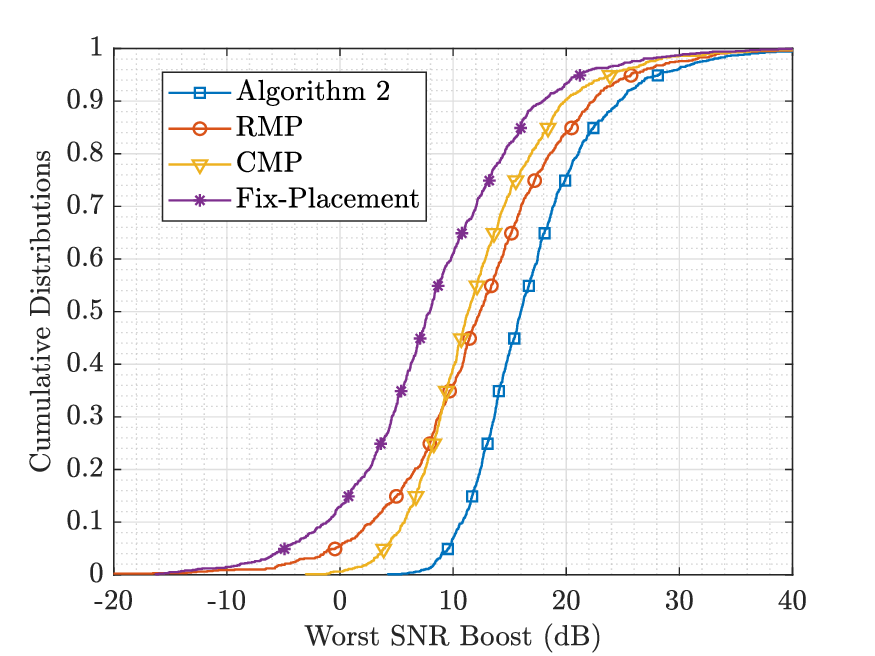}
    \caption{CDF of the worst SNR boost in the NLoS case when $U=10$.}
    \label{fig:NLOS_MU}
\end{minipage}

\begin{minipage}{0.49\linewidth}
    \centering
    \includegraphics[width=\linewidth]{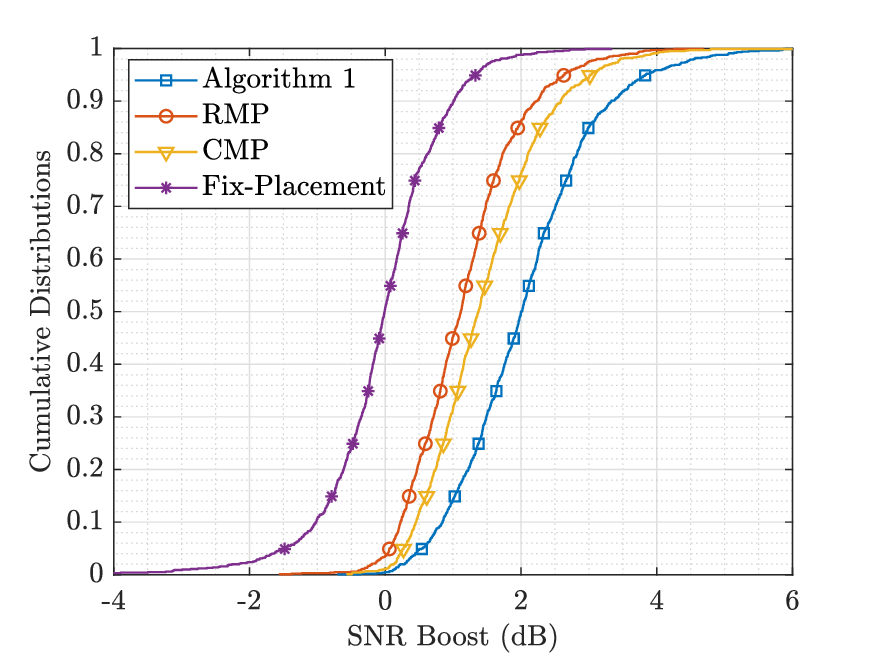}
    \caption{CDF of the SNR boost under imperfect CSI.}
    \label{fig:robust_SU}
\end{minipage}
\begin{minipage}{0.49\linewidth}
    \centering
    \includegraphics[width=\linewidth]{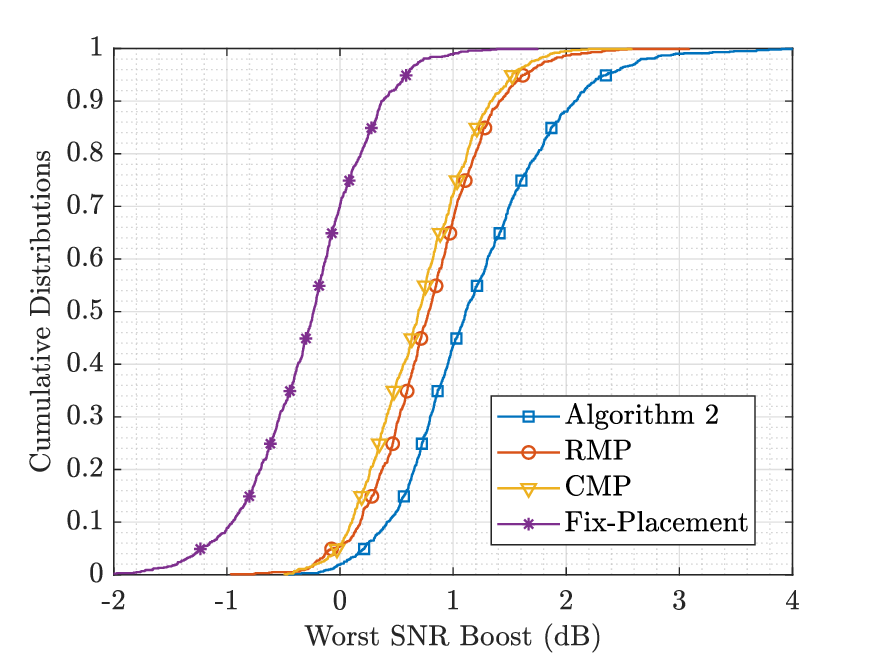}
    \caption{CDF of the worst SNR boost under imperfect CSI when $U=10$.}
    \label{fig:robust_MU}
\end{minipage}
\end{figure*}

\begin{figure}[t]
    \centering
    \includegraphics[width=\linewidth]{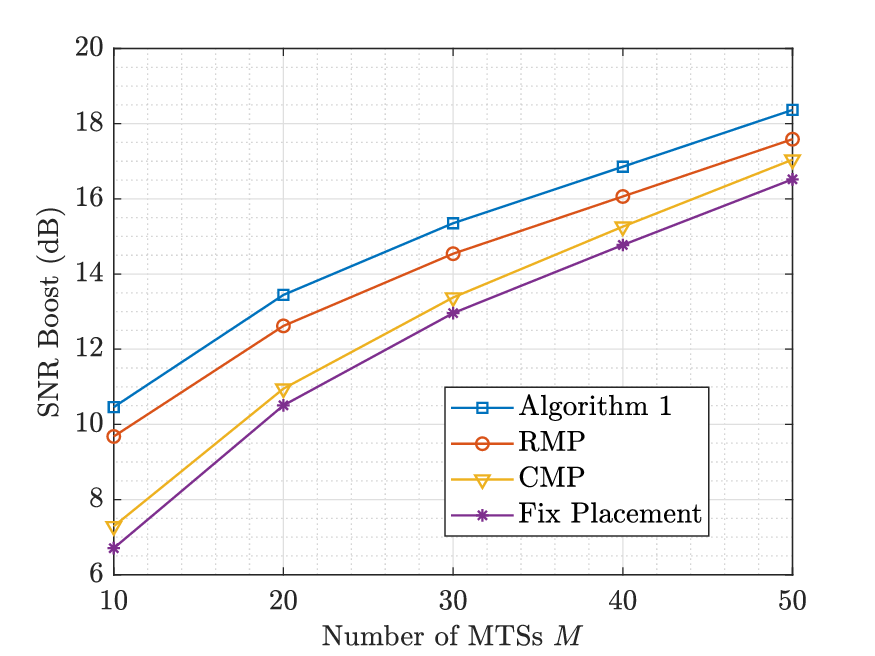}
    \caption{SNR boost vs. $M$ for MTS with phase shifters.}
    \label{fig:SNR_vs_M_phase}
\end{figure}

Let us first look at the single-receiver case. Fig.~\ref{fig:SNR_vs_M} shows the SNR boost achieved by different algorithms versus the number of MTSs $M$ when $L$ and $N$ are fixed. It can be seen that the SNR boosts achieved by the proposed Algorithm \ref{alg:OP}, the CMP, the RMP, and the \emph{Fix-Placement} method increase with $M$. This is because we can coordinate more reflection paths to enhance signal reception. It can also be seen that when $M=10$, the performance of RMP is slightly better than CMP. However, when $M$ exceeds 20, the performance of RMP becomes worse than CMP. This is because the number of possible solutions grows exponentially with $M$, whereas the number of random samples used by the RMP increases only linearly with $M$. Notice that the proposed Algorithm \ref{alg:OP} outperforms the CMP and RMP significantly. For instance, Algorithm \ref{alg:OP} improves upon the CMP by more than $1$ dB and improves upon RMP by more than $1.5$ dB when $M=50$. In contrast, the \emph{Fix-Placement} method can only achieve no more than $0.2$ dB SNR boost compared to the system without MTSs. This indicates that merely deploying movable MTSs without optimizing their placements can only bring a negligible performance improvement.

Fig.~\ref{fig:SNR_vs_L} further compares the performance of the proposed method across different grid resolutions, as indicated by varying values of $L$. It can be seen that increasing $L$ can also lead to performance improvement of different algorithms (except the \emph{Fix-Placement} method). Observe that, as $L$ increases, the performance of Algorithm \ref{alg:OP} and the CMP method first improves rapidly and then gradually stabilizes. This trend indicates that the performance gain from further increasing the number of candidate positions is limited. Observe also that, when $L=2$, the performance of the RMP is much better than that of the CMP. But when $L \geq 4$, the performance of RMP is worse than CMP. We also observe that the proposed Algorithm \ref{alg:OP} always achieves the best performance regardless of the values of $L$.

Fig.~\ref{fig:SNR_vs_N} further demonstrates how the SNR boosts achieved by different algorithms change with the number of meta-atoms $N$ of each MTS when $M$ and $L$ are fixed. It can be seen that increasing $N$ can typically bring a higher SNR boost. Moreover, as $N$ gradually increases, the performance gap between the proposed Algorithm \ref{alg:OP} and the CMP (as well as the RMP) also widens. Surprisingly, increasing $N$ does not improve the performance of the \emph{Fix-Placement} method. This suggests that simply increasing the number of reflected channels introduced by each MTS, without aligning the channels through position adjustment, does not necessarily lead to performance improvement.

We further compare the different algorithms in a broadcast network with multiple actuators. Fig.~\ref{fig:SNR_vs_U} shows the worst SNR boost achieved by different algorithms versus the number of actuators $U$. Observe that the worst SNR boost achieved by the proposed Algorithm \ref{alg:MV}, the CMP method, and the RMP method all decrease with $U$. This implies that it is increasingly difficult to coordinate the placement of multiple MTSs when more actuators are in the network. Observe also that fixing the placement of all MTSs can even lead to negative boosts. Furthermore, in the multi-receiver case, the performance of RMP and CMP is nearly identical, and the proposed Algorithm~\ref{alg:MV} achieves an additional gain of more than 1.5 dB over them.





It is worthwhile to further evaluate the performance of different algorithms in the non-line-of-sight (NLoS) transmission case, i.e., $\delta = 0$ for the direct channel $h_0$. Furthermore, the NLoS link typically experiences a greater pathloss, so the pathloss factor $\gamma$ of the direct channel is now generated according to
\begin{equation}
    \gamma= 10^{-(32.6+36.7\log_{10}(d))/10}.
\end{equation}
When $U=1$, the cumulative distributions (CDFs) of the SNR boosts achieved by the different algorithms are given in Fig.~\ref{fig:NLOS_SU}. It can be seen that, due to the weak direct path, all algorithms provide a significant SNR boost.
It can also be seen that the proposed Algorithm \ref{alg:OP} can still outperform the existing benchmarks significantly. For instance, it improves upon the CMP and RMP methods by about $4$ dB and upon the \emph{Fix-Placement} method by about $14$ dB at the $50$th percentile. Fig.~\ref{fig:NLOS_MU} further shows the worst SNR boost achieved by different algorithms in the NLoS case when $U=10$. Observe that the proposed Algorithm \ref{alg:MV} achieves the best performance.

We are also interested in the robustness of the proposed methods under imperfect CSI. We simulate the imperfect CSI by adding complex Gaussian noise with a variance of $10^{-10}$. Specifically, the imperfect CSI is used to optimize the placement, and the algorithm's performance is evaluated under exact CSI. Fig.~\ref{fig:robust_SU} and Fig.~\ref{fig:robust_MU} show the performance of different algorithms under imperfect CSI in the single-receiver and multi-receiver cases, respectively. We find that although the performance of all algorithms degrades compared to the perfect CSI case, the proposed Algorithm \ref{alg:OP} and Algorithm \ref{alg:MV} still outperform the benchmarks.

We now look at the application of the Algorithm \ref{alg:OP} in the case where MTSs are equipped with phase shifters. In particular, each meta-atom now has four discrete phase shift options. For each algorithm, we first optimize the phase shifts using the method in \cite{zhang2022configuring}, and then optimize the MTSs' placement. Fig.~\ref{fig:SNR_vs_M_phase} shows the SNR boost achieved by different algorithms versus the number of MTSs $M$. Observe that compared to the phase-shift-free MTS, using MTSs with phase shifters can further enhance the SNR boost. This is because each meta-atom can now individually adjust the phase of the reflected channel, thereby enhancing the capability to reconfigure the wireless environment. Observe also that the proposed Algorithm \ref{alg:OP} still outperforms the other methods. This indicates that even for MTSs with phase shifters, better placement can lead to further performance improvements.

Finally, we compare the running time of different algorithms in both the single-receiver and the multi-receiver cases as shown in Table \ref{tab:time}. It can be seen that the runtimes of the proposed Algorithm \ref{alg:OP} and Algorithm \ref{alg:MV} are slightly higher than that of the CMP method but lower than that of the RMP method. We also compared the runtime of the Algorithm \ref{alg:OP} with exhaustive search on a toy model with $M=10$, $L=6$, and $U=1$. Our Algorithm \ref{alg:OP} completes in $0.0023$ seconds, whereas exhaustive search takes $13.0053$ seconds. These results suggest that the proposed algorithm has low computational complexity, so it is well-suited for implementation on resource-constrained devices.

\begin{table}
\small
\renewcommand{\arraystretch}{1.3}
\centering
\caption{\small Runtimes (in Second) of different algorithms}
\begin{tabular}{|l|rr|}
\firsthline
\multicolumn{1}{|c|}{}       & \multicolumn{2}{c|}{Running time (second)}\\ \hline
\multicolumn{1}{|l|}{Algorithm} & \multicolumn{1}{r}{$U=1$} & \multicolumn{1}{r|}{$U=10$} \\ \hline\hline
Proposed Algorithm &0.0089 &0.0929 \\
CMP \cite{wang2020dynamic}& 0.0062 & 0.0694 \\
RMP \cite{you2020fast}   & 0.0157 & 0.1347 \\
\lasthline
\end{tabular}
\label{tab:time}
\end{table}

\section{Conclusion}
\label{sec:conclusion}

This paper proposes a new paradigm called \emph{FollowSpot} that enhances wireless communications by manipulating the positions of MTS on the ceiling, without requiring phase shifters at the MTS end. The optimal placement of multiple MTSs can be modeled as a nonlinear discrete optimization problem with an exponentially large solution space. It is somewhat surprising that this difficult problem can be optimally solved in polynomial time. The proposed algorithm relies on the discrete variable decoupling and the conversion of a nonconvex continuous problem to a linear search problem. Moreover, the runtime of the proposed algorithm is comparable to those of the existing heuristic methods according to simulation results.

\bibliographystyle{IEEEtran}
\bibliography{IEEEabrv,Ref}

\begin{IEEEbiographynophoto}{Wenhai Lai}(Graduate Student Member, IEEE) received the B.E. degree in information engineering from Beijing University of Posts and Telecommunications in 2021. He is currently working toward the Ph.D. degree with the School of Science and Engineering, The Chinese University of Hong Kong, Shenzhen, China. His research interests include intelligent reflecting surfaces and reinforcement learning.
\end{IEEEbiographynophoto}

\begin{IEEEbiographynophoto}{Kaiming Shen} (Senior Member, IEEE) received the B.Eng. degree in information security and the B.Sc. degree in mathematics from Shanghai Jiao Tong University, China in 2011, and then the Ph.D. degree in electrical and computer engineering from the University of Toronto, Canada in 2020. He has been with the School of Science and Engineering at The Chinese University of Hong Kong, Shenzhen, China as a tenure-track assistant professor since 2020. His research interests include optimization, wireless communications, and information theory. He currently serves as an Editor for IEEE Transactions on Wireless Communications. He is a member of the Signal Processing for Communications and Networking Technical Committee of the IEEE Signal Processing Society. He received the IEEE Signal Processing Society Young Author Best Paper Award in 2021, the University Teaching Achievement Award in 2023, the Frontiers of Science Award in 2024, and the Chinese Information Theory Society Young Researcher Award in 2025.
\end{IEEEbiographynophoto}

\begin{IEEEbiographynophoto}{Rui Zhang} (S'00-M'07-SM'15-F'17) received the B.Eng. (first-class Hons.) and M.Eng. degrees from the National University of Singapore, Singapore, and the Ph.D. degree from the Stanford University, Stanford, CA, USA, all in electrical engineering.
 
From 2007 to 2009, he worked as a research scientist at the Institute for Infocomm Research, ASTAR, Singapore. In 2010, he joined the Department of Electrical and Computer Engineering of National University of Singapore, where he is now a Provost’s Chair Professor. He is also an Adjunct Professor with the School of Science and Engineering, The Chinese University of Hong Kong, Shenzhen, China. He has published over 600 papers, all in the field of wireless communications and networks. He has been listed as a Highly Cited Researcher by Thomson Reuters/Clarivate Analytics since 2015. His current research interests include intelligent surfaces, reconfigurable antennas, radio mapping, non-terrestrial communications, wireless power transfer, AI and optimization methods.     
 
He was the recipient of the 6th IEEE Communications Society Asia-Pacific Region Best Young Researcher Award in 2011, the Young Researcher Award of National University of Singapore in 2015, the Wireless Communications Technical Committee Recognition Award in 2020, the IEEE Signal Processing and Computing for Communications (SPCC) Technical Recognition Award in 2021, the IEEE Communications Society Technical Committee on Cognitive Networks (TCCN) Recognition Award in 2023, and the IEEE James Evans Avant Garde Award in 2025. His works received 18 IEEE Best Journal Paper Awards, including the IEEE Marconi Prize Paper Award in Wireless Communications in 2015 and 2020, the IEEE Signal Processing Society Best Paper Award in 2016, the IEEE Communications Society Heinrich Hertz Prize Paper Award in 2017, 2020 and 2022, the IEEE Communications Society Stephen O. Rice Prize in 2021, etc. He served for over 30 international conferences as the TPC co-chair or an organizing committee member. He was an elected member of the IEEE Signal Processing Society SPCOM Technical Committee from 2012 to 2017 and SAM Technical Committee from 2013 to 2015. He served as the Vice Chair of the IEEE Communications Society Asia-Pacific Board Technical Affairs Committee from 2014 to 2015, a member of the Steering Committee of the IEEE Wireless Communications Letters from 2018 to 2021, a member of the IEEE Communications Society Wireless Communications Technical Committee (WTC) Award Committee from 2023 to 2025. He was a Distinguished Lecturer of IEEE Signal Processing Society and IEEE Communications Society from 2019 to 2020. He served as an Editor for several IEEE journals, including the IEEE TRANSACTIONS ON WIRELESS COMMUNICATIONS from 2012 to 2016, the IEEE JOURNAL ON SELECTED AREAS IN COMMUNICATIONS: Green Communications and Networking Series from 2015 to 2016, the IEEE TRANSACTIONS ON SIGNAL PROCESSING from 2013 to 2017, the IEEE TRANSACTIONS ON GREEN COMMUNICATIONS AND NETWORKING from 2016 to 2020, and the IEEE TRANSACTIONS ON COMMUNICATIONS from 2017 to 2022. He now serves as an Editorial Board Member of npj Wireless Technology, and the Chair of the IEEE Communications Society Wireless Communications Technical Committee (WTC) Award Committee. He is a Fellow of the Academy of Engineering Singapore. \end{IEEEbiographynophoto}

\end{document}